\documentclass[a4paper]{amsart}
\usepackage{latexsym}
\usepackage{amsmath}
\usepackage{amssymb}
\usepackage{amsfonts}
\usepackage{hyperref}
\usepackage{color}
\usepackage{enumitem}
\definecolor{Blue}{rgb}{0.,0.,1.}
\definecolor{Red}{rgb}{1.,0.,0.}

\newcounter{smallarabics}
\newenvironment{arabicenumerate}
{\begin{list}{{\normalfont\textrm{(\arabic{smallarabics})}}}
  {\usecounter{smallarabics}\setlength{\itemindent}{0cm}
   \setlength{\leftmargin}{5ex}\setlength{\labelwidth}{4ex}
   \setlength{\topsep}{0.75\parsep}\setlength{\partopsep}{0ex}
   \setlength{\itemsep}{0ex}}}
{\end{list}}

\newcounter{smallroman}

\newcommand{\ben}{\begin{arabicenumerate}}  
\newcommand{\een}{\end{arabicenumerate}}

\newtheorem{theoreme}{Theorem }[section]
\newtheorem{proposition}[theoreme]{Proposition}
\newtheorem{lemma}[theoreme]{Lemma}
\newtheorem{definition}[theoreme]{Definition}

\newtheorem{remark}[theoreme]{Remark}
\newtheorem{example}[theoreme]{Example}
\def\beq#1\eeq{\begin{align}#1\end{align}}
\def\beqa#1\eeqa{\begin{align}#1\end{align}}
\def\bes#1\ees{\begin{split}#1 \end{split}}

\newcommand{\m}{d}
\newcommand{\suppA}{\Sp\,}
\newcommand{\whGa}{\wh\Ga}

\newcommand{\bex}{\begin{example}}
\newcommand{\eex}{\end{example}}
\def\bel{\begin{lemma}}
\def\eel{\end{lemma}}
\def\bet{\begin{theoreme}}
\def\eet{\end{theoreme}}
\def\bed{\begin{definition}}
\def\eed{\end{definition}}
\def\ber{\begin{remark}}
\def\eer{\end{remark}}

\def\slim{{\rm s-}\lim}

\def\H{{\rm H}}

\def\al{\alpha}
\def\i{{\rm i}}

\def\qed{$\Box$\medskip}
\def\Sp{ \mathrm{Sp} }

\def\bbbone{{\mathchoice {\rm 1\mskip-4mu l} {\rm 1\mskip-4mu l}
{\rm 1\mskip-4.5mu l} {\rm 1\mskip-5mu l}}}
\def\one{\bbbone}

\def\12{\frac{1}{2}}

\def\supp{{\rm supp}}

\def\e{{\mathrm e}}

\def\bep{\begin{proposition}}
\def\eep{\end{proposition}}

\newcommand{\vel}{\mathrm{Vel}}
\newcommand{\cv}{\mathrm{cv}}

\newcommand{\ttau}{s}
\newcommand{\out}{\mathrm{out}}
\newcommand{\Si}{\Sigma}
\newcommand{\aloc}{\mathrm{a}\rm{-}\mathrm{loc}}
\newcommand{\mfh}{\mathfrak{h}}
\newcommand{\h}{\fr{1}{2}}
\newcommand{\ka}{\kappa}
\newcommand{\cp}{\mathrm{c}}

\newcommand{\nat}{\mathbb{N}}
\newcommand{\ti}{\tilde}
\newcommand{\w}{\mathrm{w}}
\newcommand{\un}{\overline}

\newcommand{\De}{\Delta}

\newcommand{\mcL}{\mathcal{L}}

\newcommand{\nin}{\noindent}
\newcommand{\lan}{\langle}
\newcommand{\ran}{\rangle}

\newcommand{\pa}{\partial}

\newcommand{\wt}{\widetilde}
\newcommand{\Om}{\Omega}

\newcommand{\de}{\delta}

\newcommand{\Ga}{\Gamma}

\newcommand{\hil}{\mathcal{H}}
\newcommand{\om}{\Sigma}
\newcommand{\mfa}{\mathfrak{A}}

\newcommand{\eps}{\varepsilon}
\newcommand{\fr}[2]{\frac{#1}{#2}}

\newcommand{\be}{\beta}

\newcommand{\real}{\mathbb{R}}

\newcommand{\la}{\lambda}
\newcommand{\ov}{\overline}

\newcommand{\ga}{\gamma}
\newcommand{\non}{\nonumber}

\newcommand{\wh}{\widehat}

\def\bbbone{{\mathchoice {\rm 1\mskip-4mu l} {\rm 1\mskip-4mu l}
{\rm 1\mskip-4.5mu l} {\rm 1\mskip-5mu l}}}
\def\one{\bbbone}

\def\supp{{\rm supp}}

\def\12{\frac{1}{2}}

\usepackage{verbatim}

\DeclareFontFamily{U}{mathx}{\hyphenchar\font45}
\DeclareFontShape{U}{mathx}{m}{n}{
      <5> <6> <7> <8> <9> <10>
      <10.95> <12> <14.4> <17.28> <20.74> <24.88>
      mathx10
      }{}
\DeclareSymbolFont{mathx}{U}{mathx}{m}{n}
\DeclareFontSubstitution{U}{mathx}{m}{n}
\DeclareMathAccent{\widecheck}{0}{mathx}{"71}
\DeclareMathAccent{\wideparen}{0}{mathx}{"75}

\newlength{\dinwidth}
\newlength{\dinmargin}
\begin{document}

\title[Asymptotic observables in gapped quantum spin systems]{Asymptotic observables in gapped quantum spin systems}
\author{Wojciech Dybalski}

\address{Zentrum Mathematik, Technische Universit\"at M\"unchen,
D-85747 Garching Germany}
\email{dybalski@ma.tum.de}

\keywords{quantum spin systems, Haag--Ruelle scattering theory, Araki--Haag detectors, asymptotic completeness}
\subjclass[2010]{81U99, 46N55, 82C20}

\dedicatory{Dedicated to the memory of Rudolf Haag} 

\begin{abstract} This paper gives a  construction of certain asymptotic observables 
(Araki-Haag detectors) in ground state representations of gapped quantum spin systems.
The construction is based on general assumptions which are satisfied e.g. in 
the  Ising model in strong transverse magnetic fields. 
We do not use the method of propagation estimates, but exploit
instead  compactness of the relevant propagation observables at any fixed time.
Implications for the problem of asymptotic completeness are briefly discussed.

\end{abstract}
\maketitle

\section{Introduction}
\setcounter{equation}{0}

This paper continues the model-independent discussion of scattering
theory for gapped quantum spin systems, initiated in \cite{BDN16}.
This recent work gave a construction of wave operators for such systems along the lines of Haag-Ruelle  theory \cite{Ha58, Ru62}. In the present paper we consider another classical problem of scattering theory, namely the existence of asymptotic observables. This analysis is a step towards asymptotic completeness for lattice systems, which appears to be an open problem beyond the two body scattering \cite{GS97, AB01}. We  build on recent advances in algebraic QFT \cite{DG12, DG13} but in contrast to these two references we will not use the method of propagation estimates \cite{SiSo87} to prove the existence of asymptotic observables.  We introduce a different technique, explained in more detailed below, which is based on compactness of the relevant propagation observables at any fixed time.
The underlying physical picture resembles the  Haag-Swieca compactness condition \cite{HS65},
but the mathematical implementation is quite different and does not require any additional phase space assumptions. 
The method works equally well for relativistic (algebraic) QFT and for quantum spin systems, and gives more satisfactory results than \cite{DG13}.  
In the case of quantum spin systems additional technical problems arise, in particular involving the Weyl calculus on a lattice,
which we treat in  Appendix~\ref{pseudodifferential}.

As we will discuss relativistic  QFT and lattice systems in parallel in this introduction, the group of space translations $\Ga$ will stand for $\real^d$ or $\mathbb Z^d$ here and $\wh\Ga$ will denote the Pontryagin dual.
We consider a $C^*$-dynamical system $(\mfa, \tau)$, where $\tau$ is a representation of the group of  spacetime translations $\real\times \Ga$  in automorphisms of the $C^*$-algebra of observables $\mfa\subset B(\hil)$.   
Suppose that the system is in a positive energy representation,  that is $\tau$ is implemented by a group of unitaries $U$ and 
the energy-momentum spectrum $\Sp\,U$ is contained in $\real_+\times \wh\Ga$. Suppose furthermore that $\mfa$ contains
a norm-dense subalgebra $\mfa_{\aloc}$ of `almost local' observables (see Section~\ref{Framework}).  An important insight, which
emerged in the context of relativistic QFT, is that scattering theory can be studied (in principle) in such general situation \cite{AH67,BPS91, Bu90}. The central concept is the mathematical model of a particle detector, introduced by Araki and 
Haag in \cite{AH67}
\beqa
C_t:=\int_{\Ga} d\mu(x)\, h(x/t)\tau_{(t,x)}(B^*B), \label{AH-detector}
\eeqa
where $d\mu$ is the Haar measure on $\Ga$, the observable $B\in \mfa_{\aloc}$ is `energy decreasing' (see Section~\ref{Preliminaries}) and the function $h$ is supported near the velocity  of the particle in question. The integration over whole space  $\Ga$ in (\ref{AH-detector}), which  compensates for spreading of the wave packet of the particle, allows for non-zero response of such detectors in the limit $t\to\infty$.
 Further, Araki and Haag considered coincidence arrangements of such detectors on vectors $\Psi\in \hil$
\beqa
\mathcal{C}^{\out}_n:=
\lim_{t\to\infty}\lan \Psi, C_{1,t}\ldots C_{n,t}\Psi\ran,
\label{C-n-limit}
\eeqa 
where the respective velocity functions $h_i$ have disjoint supports, and argued that the knowledge of such quantities (and their modifications accounting also for incoming particles) suffices to compute collision cross-sections. However, the difficult and still largely open part of this program is to control  the limit in (\ref{C-n-limit}).
This problem is solved in \cite{AH67} under two assumptions: First, $\Sp\, U$ contains an isolated eigenvalue at $\{0\}$ corresponding to the ground state vector $\Om$ and and isolated mass shell $\mfh$ of single-particle states (see Section~\ref{Framework}). Second,
$\Psi$  is a Haag-Ruelle scattering state of particles from $\mfh$.
Due to this second restriction, the analysis of \cite{AH67} merely corroborates the conventional scattering theory based on wave
operators and the scattering matrix.

Some progress on this problem of convergence occured only recently in \cite{DG12, DG13} in the context of algebraic QFT.  
While in these two references the
assumptions on $\Sp\, U$ are the same as in \cite{AH67}, the vector $\Psi$ does not have to be a Haag-Ruelle scattering state.
More precisely, given a small  bounded region $\De\subset \Sp\, U\backslash\{0\}$,  and choosing
$B_i^*$ to be `creation operators' of particles from $\mfh$ whose energies-momenta sum up to a point in $\De$, the limit    
\beqa
\mathcal{Q}_n^{\out}(\De)\Psi=\lim_{t\to\infty} C_{1,t}\ldots C_{n,t}\Psi 
\label{weak-AC}
\eeqa
exists for all  $\Psi$ in the spectral subspace of $\De$. Interestingly, $\mathcal{Q}_n^{\out}(\De)\Psi$ is always a Haag-Ruelle scattering
state of particles from $\mfh$ (even if $\Psi$ is not) and any scattering state from the spectral subspace of $\De$ can be obtained by this construction. Thus apart from generalizing the discussion of collision cross-sections from \cite{AH67}, this result provides a weak variant
of asymptotic completeness.
Formulated as a relation between asymptotic observables  and scattering states, it seems to be sufficient for physical interpretation of experimental data.  In fact, between a physical state $\Psi$ and a configuration of particles 
$\mathcal{Q}_n^{\out}(\De)\Psi$
there is always some intervening apparatus. Furthermore, it appears difficult to get much closer to asymptotic completeness than  (\ref{weak-AC}), without complete information about physical representations (sectors) of $\mfa$:  In fact, suppose  that $\mfa$ has some `charged' representations, disjoint from the 
vacuum representation, which also carry single-particle states in their energy-momentum spectra. Then a configuration of several such particles with total charge zero gives a state $\Psi$
in the ground state representation.  Such $\Psi$ is orthogonal to all Haag-Ruelle scattering states of particles from $\mfh$ and thus  the conventional asymptotic completeness fails in this situation. But the weaker concept discussed above remains valid, as such $\Psi$ is simply
annihilated by the detectors in (\ref{weak-AC}).

In this paper we establish (\ref{weak-AC}) for gapped quantum spin systems, using a novel method for controlling the convergence of asymptotic observables, which we will now explain.
For the purpose of demonstration we consider  the quantum-mechanical Hamiltonians $H=\h p^2+V(x)$ and $H_0=\h p^2$ on $L^2(\real^d)$, where $p=-\i\nabla_x$ and $V$ is some rapidly decreasing repulsive potential. The problem of existence of the wave
operator 
\beqa
W^{\out}=\lim_{t\to\infty}\e^{\i tH} \e^{-\i tH_0}
\eeqa 
is readily solved by applying the Cook's method to the sequence $W_t:=\e^{\i t H} \e^{-\i t H_0}$, i.e.
noting that $t\to \|\pa_t W_t\psi\|=\|V(x)\e^{-\i t H_0}\psi\|$ is integrable in $t$ on some dense domain of
vectors $\psi\in L^2(\real^d)$. A toy model of an Araki-Haag detector  has the form
\beqa
c_t:=\e^{\i tH}\chi(p) h(x/t)\chi(p)\e^{-\i tH},
\eeqa    
where $\chi, h\in C_0^{\infty}(\real^d)$, $0\notin \supp\, h$. (For technical reasons specific to our method we also require  that $\supp\chi\subset \supp\,h$ and $\supp\, h$ is a convex set).
Here an attempted application of the Cook method  gives
\beqa
\pa_t c_t=\fr{1}{t}\e^{\i tH}\chi(p)(\nabla_x h)(x/t)\big(p-x/t\big)\chi(p)\e^{-\i tH}+O(t^{-2}),
\eeqa
which is not manifestly integrable. Certainly in this simple situation we could proceed via the method of propagation estimates
which (roughly speaking) amounts to guessing a suitable family of auxiliary sequences $c'_t$, computing $\pa_t c'_t$ and eliminating  
the problematic terms from the resulting system of inequalities. The
underlying physical mechanism is that the difference between the instantaneous and average velocity  $(p-x/t)$ tends to zero along the asymptotic, ballistic trajectory of the particle \cite{Gr90}.
 
However, it  turns out that for our particular asymptotic observable there is a more direct way to proceed. It starts from a simple observation
that for $\psi\in \mathrm{Ran}\, W^{\out}$ the limit $\lim_{t\to\infty} c_t\psi$ is readily computed, thus it suffices
to consider $\psi\in (\mathrm{Ran}\, W^{\out})^{\bot}$. For such vectors we can write
\beq\bes \label{new-argument}
c_t\psi&=\e^{\i tH}\e^{-\i tH_0}\chi(p) h(x/t+p)\chi(p)\e^{\i tH_0}\e^{-\i tH}\psi\\
&=\e^{\i tH}\e^{-\i tH_0}\chi(p) h(x/t+p)\chi(p) (W_t^*-(W^{\out})^*)\psi \\
&=\e^{\i tH}\e^{-\i tH_0}\chi(p) h(x/t+p)\chi(p) \int_t^{\infty}ds\, (-\pa_{s}W^*_{s})\psi \\
&=\e^{\i tH}\e^{-\i tH_0}\chi(p) h(x/t+p)\chi(p) \int_t^{\infty}ds\, \i\e^{\i s H_0}V(x)\e^{-\i s H}\psi.
\ees\eeq
Here in the second step we used that $(W^{\out})^*\psi=0$, in the third step we exploited compactness of $\chi(p) h(x/t+p)\chi(p)$
and in the last step we obtained an expression which tends to zero in norm. This follows from the decay of the potential, the assumptions on the supports of $\chi, h$ and the equality $g(x/t)\leq g(x/s)$ valid for
$s\geq t$ if $g$ is the characteristic function of a convex set containing zero.  Some similarity  with  existing methods of scattering theory has to be admitted (see the proofs of Theorems XI.7 and XI.112 of \cite{RS3}),
nevertheless the strategy of controlling asymptotic observables presented in (\ref{new-argument}) seems to be new.  

Of course in the above quantum mechanical situation the method does not give any new information, since asymptotic completeness of such  Hamiltonians is well known from the outset. But for systems of $n>2$
interacting particles with non-quadratic dispersion relations, as for example relativistic QFT or spin systems, it has advantages. In such 
systems only observables which keep the average and instantaneous velocity of a particle very close together could (so far) be handled by the method of propagation estimates  \cite{DG13}.
The alternative method explained above only requires  that the set of average velocities contains the set of instantaneous velocities (see Theorem~\ref{Main-result}  below for precise assumptions).
To cover the case when the two sets are disjoint it seems necessary to combine the method with Mourre theory, which is so far an open problem. Another important open problem, which probably requires a different approach, is to control convergence  (or  existence of non-zero limit points) of Araki-Haag detectors without assuming mass shells in the energy-momentum spectrum.

This paper is organized as follows:  Sections~\ref{Framework} and \ref{Preliminaries} are devoted to our framework and preliminaries.
In Section~\ref{Main} we state and prove our main result. Some auxiliary results  are postponed to Appendix~\ref{AppendixA}. Appendix~\ref{pseudodifferential} concerns pseudo-differential calculus on a lattice.  

\vspace{0.5cm}

\nin\textbf{Acknowledgements:} I would like to thank C. G\'erard, 
S. Bachmann and P. Naaijkens for earlier collaborations which 
provided a basis for this work. Furthermore, I acknowledge interesting discussions on the subject of this paper with G.M. Graf and J.S. M\o ller. This work was  supported by the DFG within the Emmy Noether grant DY107/2-1.

\section{Framework} \label{Framework}
\setcounter{equation}{0}

We work in a framework  outlined in the introduction of \cite{BDN16} which is suitable for gapped quantum spin systems in 
irreducible ground state representations.
Let $\Ga=\mathbb{Z}^d$ be the abelian group of space translations and $\wh\Ga$ its Pontryagin dual which is in our case
$S_1^d$ i.e. the Brillouin zone.  We will often  use  the parametrisation $S_1^d=]-\pi,\pi]^{d}$. Concerning the Schwartz classes $S(\real\times \Ga), S(\Ga)$  and Fourier transforms we follow the conventions and notation of Appendix D of \cite{BDN16}.

We consider a $C^*$-dynamical system $(\mfa, \tau)$, where $\tau$ is a 
strongly continuous representation of space-time translations $\real\times \Ga$ in automorphisms of a concrete $C^*$-algebra
$\mfa$, which acts irreducibly on a Hilbert space $\hil$.  We denote
\beqa\bes\label{smearing-intro}
&\tau_{f}(A):= ( 2 \pi)^{-\frac{d+1}{2}} \int_{\real\times\Ga} dt d\mu(x)\, \tau_{(t,x)}(A) f(t,x), \quad f\in L^1(\real\times \Ga), \\
&\tau^{(d)}_{g}(A):= (2 \pi)^{-\frac{d}{2}} \int_{\Ga}  d\mu(x)\, \tau_{x}(A) g(x), \quad\quad\quad\quad\quad \ g\in L^1(\Ga), 
\ees\eeqa
where $\mu(\De):=\sum_{x\in \De} 1$  is the Haar mesure on $\Ga$. We require that $\mfa$ contains a norm-dense $*$-subalgebra of \emph{almost local} observables
$\mfa_{\aloc}\subset \mfa$ s.t. for any $A, A_1, A_2\in\mfa_{\aloc}$
\beqa\bes\label{almost-local-intro}
 &\tau_{(t,x)}(A) \in \mfa_{\aloc},\quad (t,x)\in\real\times \Ga, \\
&\tau_{f}(A)\in \mfa_{\aloc}, \quad\quad\quad f\in S(\real\times \Ga), \\ 
&\tau_{g}^{(d)}(A)\in  \mfa_{\aloc}, \quad\quad \ g\in S(\Ga),  \\ 
&[\tau_{x_1}(A_1), \tau_{x_2}(A_2)]=O( \lan x_1-x_2\ran^{-\infty}).
\ees\eeqa
The existence of such subalgebra 
$\mfa_{\aloc}$ in gapped quantum spin systems was shown in \cite{BDN16, Sch83} with the help of the Lieb-Robinson bounds.

Furthermore, we assume that $\tau$ is implemented by a strongly continuous group of unitaries $U$,~i.e.
\beqa
\tau_{(t,x)}(A)=U(t,x)AU(t,x)^*, \quad A\in \mfa.
\eeqa
We denote by $P(\,\cdot\,)$ the spectral  measure of $U$ given by the SNAG theorem and denote its support by $\Sp\, U\subset \real\times \wh\Ga$.
We impose several restrictions on this energy-momentum spectrum:
\begin{enumerate}
\item \textbf{Positivity of energy.} $\Sp\, U\subset \real_{+}\times \wh\Ga$.
\item \textbf{Ground state vector.} $(0,0)$ belongs to  $\Sp\, U$ and is an isolated, simple eigenvalue corresponding to the eigenvector $\Om$.
\item \textbf{Single-particle states.} There is an isolated mass-shell $\mfh\subset \Sp\, U$ which is a graph of a function 
$\om\in C^{\infty}(\wh\Ga)$ whose Hessian matrix vanishes at most on a subset of Lebesgue measure zero. Moreover, $(\mfh-\mfh)\cap \Sp\, U=\emptyset$.
\end{enumerate}
As indicated in \cite{BDN16}, all our assumptions hold in the Ising model in strong transverse magnetic fields in any space dimension.

\section{Preliminaries}\label{Preliminaries}
\setcounter{equation}{0}

The Arveson spectrum (or the energy-momentum transfer) of an observable $B\in \mfa$ is denoted $\Sp_B\tau$.
We recall that $(E,p)\in \Sp_B\tau$ if for any neighbourhood $V$ of this point there is $f\in L^1(\real\times \Ga)$,
whose Fourier transform $\wh f$ is supported in $V$ and $\tau_{f}(A)\neq 0$. There holds 
\beqa\bes
& \Sp_{B^*}\tau=-\Sp_{B}\tau, \\
& BP(\De)\hil\subset P(\ov{\De+\Sp_{B}\tau})\hil,
\ees\eeqa
where $\De\subset \real\times \wh\Ga$ is a Borel subset.  Clearly, for $A\in \mfa_{\aloc}$ and $f\in S(\real\times \Ga)$
we have $\tau_f(A)\in\mfa_{\aloc}$ and $\Sp_{\tau_f(A)}\tau\subset \supp\, \wh f$.  For a comprehensive discussion of the Arveson spectrum
in the context of spin systems we refer to \cite{BDN16}.

Now we recall the Haag-Ruelle scattering theory for spin systems developed in  \cite{BDN16}: 
Let $B_i^*\in \mfa_{\aloc}$, $i=1,\ldots, n$, be s.t. $\Sp_{B_i^*}\tau\cap \Sp\, U\subset \mfh$ and consider positive-energy
wave packets of particles from $\mfh$
\beqa
g_{i,t}(x):=(2\pi)^{-\fr{d}{2}}\int_{\wh\Ga} dp\, \e^{-\i\Si(p)t+\i p\cdot x} \wh{g}_{i}(p), \quad \wh{g}_i\in C^{\infty}(\wh\Ga).\label{wave-packet-def}
\eeqa
The set $\vel(\supp\,\wh{g}):=\{\nabla \Si(p)\,|\, p\in \supp\,\wh{g}\,\}$
is called the velocity support of $\supp\,\wh{g}$. 
The Haag-Ruelle creation operators are defined as
\beqa
B^*_{i,t}(g_{i,t}) := (2 \pi)^{-\frac{d}{2}} \int_{\Ga} d\mu(x)\, B_{i,t}^*(x)g_{i,t}(x),
\eeqa
where $B_{i,t}^*(x):=B_i^*(t,x):=U(t,x)B_i^*U(t,x)^*$.
\bet\cite{BDN16} Let $B_i^*$, $\wh {g}_{i}$  be as above and s.t. $\supp\,\wh{g}_i$ have mutually disjoint velocity supports. 
Then there exist $n$-particle scattering states given by
\beqa
\Psi^{\out}=\lim_{t\to\infty} B^{*}_{1,t}(g_{1,t})\ldots B^{*}_{n,t}(g_{n,t})\Om. \label{scattering-state-one}
\eeqa
Moreover,  such states for $n=0,1,2\ldots$ span a subspace $\hil^{\out}\subset \hil$ which is naturally isomorphic to
the symmetric Fock space over the single-particle space $\hil_{\mfh}:=P(\mfh)\hil$.  
\eet
\nin This theorem belongs to a long list of results 
adapting the relativistic Haag-Ruelle theory  to various non-relativistic 
models and frameworks, see e.g. \cite{BF91, Al73, NRT83}. We refer to \cite{BDN16} for a more thorough discussion and comparison
with the literature.

Next, we denote by $\mcL_0$ the subspace of all observables which are almost local and energy decreasing i.e. $B\in \mfa_{\aloc}$ and
$\Sp_B\tau\subset (-\infty,0)\times \whGa$. For such operators there holds an important result from \cite{Bu90} which 
was adapted to the lattice case in \cite{BDN16}:
\bet\label{harmonic-theorem} Let $\De\subset \real\times \wh\Ga$ be  compact and $B\in \mcL_0$.  
Then, for any finite set $K\subset\Ga$,
\beqa
\|P(\De)\int_{K} d\mu(x)\, (B^*B)(x) P(\De)\|\leq c_{\De}, \label{B-bound}
\eeqa
where $c_{\De}$ is independent of $K$.
\eet
\nin In view of Theorem~\ref{harmonic-theorem}, the following maps are well defined 
for any $B_1,\ldots, B_n\in \mcL_0$:
\beqa\bes
&a_{B_n,\ldots, B_1}:  \hil_{\cp} \to \hil \otimes L^2(\Ga^{n}),\\
& (a_{B_n,\ldots, B_1}\Psi)(x_1,\ldots, x_n)=B_n(x_n)\ldots B_1(x_1)\Psi,
\ees\eeqa  
where $\hil_{\cp}$ is the domain of vectors of bounded energy, i.e. such $\Psi\in \hil$ that
$\Psi=P(\De)\Psi$ for some compact $\De$.
For brevity we will write $a_{\un{B}}:=a_{B_n,\ldots, B_1}$.

The maps $a_{\un{B}}$ where introduced and thoroughly studied in \cite{DG12,DG13} in the context
of local relativistic QFT and this discussion is easy to adapt to the lattice framework. Here we merely
point out that $a_{\un{B}}P(\De): \hil \to \hil \otimes L^2(\Ga^n )$  is a bounded map for any compact 
set $\De$  and therefore
\beqa
\|P(\De)a_{\un{B}}^*(A\otimes b)a_{\un{B}} P(\De)\|\leq c_{\De}\|A\|\|b\|, 
\label{key-bound}
\eeqa
where $A\in B(\hil)$, $b\in B(L^2(\Ga^n))$ and $c_{\De}$ is independent of $A, b$.

\section{Main result}\label{Main}
\setcounter{equation}{0}

For $h\in C_0^{\infty}(\real^d)$,  $B\in\mcL_0$ we define the approximating sequence of an Araki-Haag detector:
\beqa
C_t:=\int_{\Ga} d\mu(x)\, h(x/t)(B^*B)(t,x)=a_{B_t}^*(1\otimes h_t)a_{B_t}, \label{one-detector}
\eeqa
where the second equality holds on $\hil_{\cp}$ and $h_t(x):=h(x/t)$ is understood as a multiplication operator on $L^2(\Ga)$.
Our main result, stated in Theorem~\ref{Main-result} below, concerns the strong convergence  as $t\to\infty$  of 
products of such operators $C_{i,t}$ on the range of $P(\De)$,
where $\De\subset \real\times \wh\Ga$ is an open bounded set. 
Similarly as in \cite{DG13}, the set $\De$ and Arveson spectra of $B_i$ are linked by the following  condition:
\begin{definition}\label{delta-admissible}
 Let $\Delta\subset \real\times \wh{\Ga}$ be an open bounded set 
 and $B_{1},\ldots, B_{n}\in \mcL_{0}$. We say that $\un B=(B_{n},\ldots, B_{1})$ is $\Delta-${\em admissible} if 
 \beq\label{transfer-to-hyperboloid}
& \Sp_{B^*_i}\tau \cap \Sp\, U\subset \mfh,\ i=1,\ldots, n,\\[2mm]
&  \Sp_{B^*_1}\tau +\cdots+ \Sp_{B^*_n}\tau \subset \De, \label{transfer-from-vacuum}\\[2mm]
& \big(\ov{\De}-(\Sp_{B^*_1}\tau +\cdots+ \Sp_{B^*_n}\tau) \big)\cap \Sp\, U\subset \{0\}, \label{transfer-to-vacuum}
\eeq
and addition is meant in the abelian group $\real\times \wh\Ga$.
\end{definition}
This condition  essentially says that $B_i^*$ are `creation operators' of particles with energies-momenta
near some  $(E_i, p_i)\in \mfh$, which add  up to a point in $\De$. In other words, if we used these $B_i^*$ to construct the scattering
state $\Psi^{\out}$ given by (\ref{scattering-state-one}), then $\Psi^{\out}\in P(\De)\hil$.  On the other hand, if for a given set $\De$
there is no $\De$-admissible $\un{B}$, then also $P(\De)\hil^{\out}=\{0\}$.

 Finally, we introduce some notation used in the following theorem and its proof: For any $X\subset \wh{\Ga}$ we define the velocity support $\vel(X):=\{ \nabla\om(p')\,|\, p'\in  X\,  \}\subset \real^d$. 
For $Y\subset \real\times \wh{\Ga}$, understood as a subset
of the energy-momentum spectrum, we write
$\pi_p(Y):=\{ p'\in \wh\Ga\,|\, (E',p')\in Y\,\}$. For $Y\subset \real^{\m}\times \wh{\Ga}$, understood as a subset of the phase space (cf.  Appendix~\ref{pseudodifferential}), we write 
$\pi_x(Y):=\{ x'\in \real^{d}\,|\, (x',\xi')\in Y\,\}$. 
The convex hull of a set $Z\subset \real^d$ will be denoted by $Z^{\cv}$.   We will denote by $Z^{\de}$ a slightly larger set than $Z$, 
namely $Z^{\de}:=Z+\mathcal{B}(0,\de)$, where    $\mathcal{B}(0,\de)$ is a closed ball of radius $\de$ centered at zero.
\bet\label{Main-result} Let $\De\subset \real\times \wh\Ga$ be an open bounded set and
$B_i\in\mcL_0$, $i=1,\ldots,n$, be $\De$-admissible  in the sense of Definition~\ref{delta-admissible} and s.t. $\pi_p(\Sp_{B^*_i}\tau)$ are disjoint sets in the interior of $]-\pi,\pi]^d$. Let $h_i\in C_0^{\infty}(\real^d)$ be s.t. 
$\supp\, h_i\supset  \vel(\pi_p(\Sp_{B^*_i}\tau))$ and
$(\supp\, h_i)^{\cv}$ are disjoint sets. Then, for $C_{i,t}$  given by (\ref{one-detector}) and any $\Psi\in P(\De)\hil$
there exists the limit
\beqa
\mathcal{Q}^{\out}_n(\De)\Psi:=\slim_{t\to\infty}C_{1,t}\ldots C_{n,t}\Psi,
\label{asymptotic-observable}
\eeqa
and belongs to $P(\De)\hil^{\out}$. Moreover, $P(\De)\hil^{\out}$ is the closed span of vectors of the form (\ref{asymptotic-observable}) for $n=1,2,3,\ldots$ provided that
 $(\ov{\De}-\ov{\De})\cap \Sp\, U=\{0\}$ and $0\notin \De$. 
\eet
\proof We first note that for $\Psi_2\in P(\De)\hil^{\out}$ one can show the existence of $\mathcal{Q}^{\out}_n(\De)\Psi_2$ 
by standard arguments \cite{AH67, DG13}, which will not be repeated here. For similar reasons we skip the proof of the last statement of the theorem, which follows closely the reasoning from Section~7 of \cite{DG13} (with some input from Subsection 4.4 of \cite{BDN16}). 
We focus here on the most essential part of the proof, which consists in showing that for
$\Psi_2\in P(\De)(\hil^\out)^{\bot}$ the sequence
\beqa
t\mapsto \lan \Psi_1,C_{1,t}\ldots C_{n,t}\Psi_2\ran, \label{equation}
\eeqa
tends to zero uniformly in  $\Psi_1\in\hil_\mathrm{c}$, $\|\Psi_1\|\leq 1$.
\newcommand{\chii}{g}
\newcommand{\etai}{\wh{g}}
Due to almost locality of $B_i$, disjointness of supports of $h_i$ and $\De$-admissibility, expression~(\ref{equation}) equals, up to a term of order $\|\Psi_1\| \|\Psi_2\| O(t^{-\infty})$,  
\beqa
I_t:=\lan \Psi_{1,t},a^*_{\un{B}}(\, |\Om\ran\lan \Om| \otimes  h_{1,t}\ldots  h_{n,t} )a_{\un{B}}\Psi_{2,t}\ran.
\label{hit}
\eeqa
Now by a partition argument, it suffices to consider
\beqa
I_t'=\lan \Psi_{1,t}, a^*_{\un{B}}(\,  |\Om\ran\lan \Om|  \otimes  h_{1,t}\ldots  h_{n,t} )a_{\un{B}^{\prime} }\Psi_{2,t}\ran,
\eeqa
where  $\Sp_{B^{\prime *}_i}\tau\subset \Sp_{B^*_i}\tau$  are sufficiently small, say contained in balls of radii  $0<\de\ll 1$.
Before we proceed, let us introduce some notation: 
\beqa\bes
\ti x&:=(x_1,\ldots, x_n),\\
\ti k&:=(k_1,\ldots, k_n),\\
 D_{\ti x}&:=(D_{x_1}, \ldots, D_{x_n}), \\ 
\ti\om(\ti k)&:=\om(k_1)+\cdots+\om(k_n), \\
\etai(\ti k)&:=\wh\chii_{1}(k_1)\ldots \wh\chii_{n}(k_n),\label{eta-def} \\
\H(\ti x)&:=h_{1}(x_1)\ldots h_{n}(x_n), \quad \H_t(\ti x):=\H(\ti x/t), \\
D&:=\{\, \ti x\in \real^{nd} \,|\, x_i=x_j \textrm{ for some } i\neq j\,\},
\ees\eeqa 
where   the lattice momentum operators $D_{x_i}$ are defined in Appendix~\ref{pseudodifferential} and $\wh g_i\in C^{\infty}(\wh\Ga)$
are equal to one on $\pi_p(\Sp_{B^{\prime *}_i}\tau)$ and supported in  slightly larger sets $(\pi_p( \Sp_{B^{\prime *}_i}\tau ))^{\de}$ which are in 
the interior of $]-\pi,\pi]^d$.
We note that by the disjointness of supports of $h_i$ we have $\supp\, \H\cap D=\emptyset$. 
Moreover, by  Lemma~\ref{momentum-cons-lemma} we can write
\beqa
I_t'=&\lan \Psi_{1,t},a^*_{\un{B}}\big(\,|\Om\ran\lan \Om|
\otimes    \H_t(\ti x ) \etai^2(D_{\ti x})  \big)a_{\un{B}'}\Psi_{2,t}\ran. 
\label{I-t-formula}
\eeqa
Now we choose $\etai'_i\in C^{\infty}(\wh\Ga)$ s.t. $\etai'_i\etai_i=\etai_i$ and $\etai'_i$ are supported in 
$(\pi_p(\suppA_{ B_i^{\prime *}} \tau  ))^{2\de}$.
Setting $\wh\chii'(D_{\ti x}):=\wh\chii'_{1}(D_{x_1})\ldots \wh\chii'_{n}(D_{x_n})$, we have by the pseudo-differential calculus (see Appendix~\ref{pseudodifferential})
\beqa\bes\label{pseudo-computation}
 \H_t(\ti x ) \etai^2(D_{\ti x})&=(\H_t\etai')^{\w}\etai(D_{\ti x})^2+O(t^{-1})\\
&=\e^{-\i t\ti\om(D_{\ti x})}(\H_{1,t}\etai')^{\w}
\etai(D_{\ti x})^2\e^{\i t\ti\om(D_{\ti x})}+O(t^{-1})\\
&=\e^{-\i t\ti\om(D_{\ti x})}(\H_{1,t}\etai')^{\w}   \etai(D_{\ti x}) \H_{0,t}(\ti x)\etai(D_{\ti x})\e^{\i t\ti\om(D_{\ti x})}+O(t^{-1}),
\ees\eeqa
where $\H_{1}(\ti x, \ti\xi)=\H(\ti x+\nabla\ti\om(\ti\xi))$ and $\H_0\in C_0^{\infty}(\real^{nd})$, $0\leq \H_0\leq 1$,
is equal to one on $\pi_{\ti x}\, \supp\, (\H_1\etai')$ and supported in
$(\pi_{\ti x}\, \supp\, (\H_1\etai') )^{\de}$. 
Decomposing $I'_t=I_t''+O(t^{-1})$ in accordance with (\ref{pseudo-computation}) and (\ref{key-bound}), and making use of Lemmas~\ref{Bostelmann}, \ref{HR-states}, we get 
\beqa\bes\label{sum}
I_t''&=\sum_{\al,\be\in \nat_{0}^{nd} }
\fr{  {\wt \H_{0,t}}^{(\al,\be)} }{\al!\be!}\lan \Psi_{1,t},a^*_{\un{B}}\big(   |\Om\ran\lan \Om|
\otimes \e^{-\i t\ti\om(D_{\ti x})} (\H_{1,t}\etai')^{\w}|\etai(\ti k)\ti k^{\al}\ran \lan \etai(\ti k)\ti k^{\be}|\e^{\i t\ti\om(D_{\ti x})}   \big)a_{\un{B}'}\Psi_{2,t}\ran\\
&=\sum_{\al,\be \in  \nat_{0}^{nd}}
\fr{  {\wt \H_{0,t}}^{(\al,\be)} }{\al!\be!}
\lan \Psi_{1,t},a^*_{\un{B}}\big(|\Om\ran\otimes \e^{-\i t\ti\om(D_{\ti x})} (\H_{1,t}\etai')^{\w}|\etai(\ti k)\ti k^{\al}\ran )(  \lan \Om|\otimes  \lan \etai(\ti k)\ti k^{\be}|\e^{\i t\ti\om(D_{\ti x})}   \big)a_{\un{B}'}\Psi_{2,t}\ran\\
&=\sum_{\al,\be \in  \nat_{0}^{nd} }
\fr{  {\wt\H_{0,t}}^{(\al,\be)}}{\al!\be!}\lan \Psi_{1,t},a^*_{\un{B}}\big(|\Om\ran\otimes \e^{-\i t\ti\om(D_{\ti x})} (\H_{1,t}\etai')^{\w}|\etai(\ti k)\ti k^{\al}\ran )
 \lan  B^{\prime*}_{1,t}(g_{1,t}^{\be_1})\ldots  B^{\prime*}_{n,t}( g_{n,t}^{\be_n})\Om, \Psi_{2}\ran,
\ees\eeqa
where $g_{i,t}^{\be_j}$ are wave packets introduced in Lemma~\ref{HR-states}. By our assumptions, $\vel(\pi_p(\Sp_{B^{\prime *}_i}\tau))$ are disjoint sets,
thus velocity supports of $\supp\, \wh g_i^{\be}$ are disjoint for $\de$ small. 
We obtain from Lemma~\ref{Bostelmann}, that 
\beqa
|\wt\H_{0,t}^{(\al,\be)}| \leq t^{ |\al|+|\be|+nd } c^{ |\al|+|\be| }, \label{x}
\eeqa
thus it may appear difficult to obtain convergence of (\ref{sum}) to zero  as $t\to\infty$. However, we recall that $\Psi_2$ is orthogonal  to all scattering
states and therefore the term
\beqa
F(t):=\lan  B^{\prime *}_{1,t}(g_{1,t}^{\be_1})\ldots  B^{\prime *}_{n,t}( g_{n,t}^{\be_n})\Om, \Psi_{2}\ran
\eeqa
tends to zero as $t\to\infty$.
Let us analyse this expression along the lines of Haag-Ruelle theory:
\beqa\bes
F(t)&=\lan B_{1,t}^{\prime *}(g_{1,t}^{\be_1}) \ldots B_{n,t}^{\prime *}(g_{n,t}^{\be_n})\Om, \Psi_2\ran\\
&=-\int_t^{\infty}d\ttau 
\lan \pa_{\ttau}(B_{1,\ttau}^{\prime *}(g_{1,\ttau}^{\be_1})\ldots  
B_{n,\ttau}^{\prime *}(g_{n,\ttau}^{\be_n}))\Om, \Psi_2\ran\\
&=-\sum_{\ell=1}^{n-1}\sum_{k=\ell+1}^n\int_t^{\infty}d\ttau 
\lan B_{1,\ttau}^{\prime *}(g_{1,\ttau}^{\be_1})\ldots 
[\pa_{\ttau}(B_{\ell,\ttau}^{\prime *}(g_{\ell,\ttau}^{\be_{\ell}})),  B_{k,\ttau}^{\prime*}(g_{k,\ttau}^{\be_{k}}) ]\ldots B_{n,\ttau}^{\prime *}(g_{n,\ttau}^{\be_n})\Om, \Psi_2\ran.
\label{commutator-gain}
\ees\eeqa
Having gained the commutator in (\ref{commutator-gain}) we will
essentially reverse the steps which led from (\ref{I-t-formula}) to (\ref{sum})
and then obtain convergence to zero using locality and the fact that $\supp\, \H\cap D=\emptyset$.
(It does not seem possible to exploit this latter property directly in 
formula~(\ref{sum})). Thus we substitute (\ref{commutator-gain}) to (\ref{sum}), exchange the order of summation w.r.t.
$\al,\be$  and integration w.r.t. $\ttau$ using Lemma~\ref{decay-of-terms}
and  sum back the series w.r.t. $\al, \be$. To avoid long formulas, we note that $\pa_{\ttau}(B_{\ell,\ttau}^{\prime *}(g_{\ell,\ttau}^{\be_{\ell}}))=
\dot{B}_{\ell,\ttau}^{\prime *}(g_{\ell,\ttau}^{\be_{\ell}})+{B}_{\ell,\ttau}^{\prime *}(
\dot{g}_{\ell,\ttau}^{\be_{\ell}})$, where dots indicate time derivatives, and denote by $I_{(\ell,k),t}''$ the contribution to $I''_t$ coming from the term with
the commutator $[\dot{B}_{\ell,\ttau}^{\prime *}(g_{\ell,\ttau}^{\be_{\ell}}),  B_{k,\ttau}^{\prime *}(g_{k,\ttau}^{\be_{k}}) ]$ in (\ref{commutator-gain}). (The analysis of  terms involving 
$\dot{g}_{\ell,\ttau}^{\be_{\ell}}$ is analogous and will be skipped here). We obtain
\beqa
I_{(\ell,k), t}'' =\int_{t}^{\infty}d\ttau \lan F_{1,t}, \e^{-\i t\ti\om(D_{\ti x})}(\H_{1,t}\etai')^{\w}
\etai(D_{\ti x}) \H_{0,t}(\ti x)\etai(D_{\ti x})\e^{\i \ttau\ti\om(D_{\ti x})} F_{2,\ttau}\ran_{L^2(\Ga^n )},
\eeqa
where we introduced two $L^2$ functions
\beqa\bes
&F_{1,t}(x_1,\ldots, x_n):=\lan \Psi_{1,t}, B_1^*(x_1)\ldots B_n^*(x_n)\Om\ran, \\
& F_{2,\ttau}(x_1,\ldots, x_n):=-\lan \Psi_{2,\ttau},B_1^{\prime *}(x_1)\ldots  [\dot{B}_\ell^{\prime*}(x_\ell), B_k^{\prime*}(x_k)]\ldots B^{\prime*}_n(x_n)\Om\ran.
\ees\eeqa
Since $\sup_t\|F_{1,t}\|_2\leq c\|\Psi_1\|$ by Theorem~\ref{harmonic-theorem}, we get
\beqa\bes
| I_{(\ell,k), t}''  |\leq C\|\Psi_1\|\int_t^{\infty} d\ttau
\|  \H_0(\ti x/t)  \e^{\i\ti\om(D_{\ti x})\ttau}\etai(D_{\ti x})F_{2,\ttau}\|_2.
\ees\eeqa
Now we recall that $0\leq \H_0\leq 1$ and $\supp\,\H_0\subset  (\pi_{\ti x}\, \supp\, (\H_1\etai') )^{\de}$.
We denote by $\chi$ the sharp characteristic function of $(\pi_{\ti x}\, \supp\, (\H_1\etai') )^{\de,\cv}$. Making use of the
fact that the latter set is convex and contains zero, (here we use the assumption that $\supp\, h_i\supset  \vel(\pi_p(\Sp_{B^*_i}\tau))$), we  estimate
\beqa
\H_0^2(\ti x/t)\leq  \chi(\ti x/t)\leq \chi(\ti x/\ttau)\leq (\H_0^{\prime})^2(\ti x/\ttau) \textrm{ for }  t\leq \ttau,
\eeqa
where $\H_0^{\prime}\in C_0^{\infty}(\real^{nd})$ is equal to one on $(\pi_{\ti x}\, \supp\, (\H_1\etai') )^{\de,\cv}$ and supported
in $(\pi_{\ti x}\, \supp\, (\H_1\etai') )^{\de,\cv,\de}$.
Thus we can write using Lemmas~\ref{trivial-pseudo}, \ref{free-evolution-x}, 
\beqa
\|  \H_0(\ti x/t)  \e^{\i\ti\om(D_{\wt{x}}) \ttau}\etai(D_{\ti x})F_{2,\ttau}\|_2&\leq 
\|  \H_0'(\ti x/\ttau)  \e^{\i\ti\om(D_{\wt{x}}) \ttau}\etai(D_{\ti x})F_{2,\ttau}\|_2\non\\
 &\leq \|  (\H_{0,\ttau}'\etai)^{\w}  \e^{\i\ti\om(D_{\wt{x}}) \ttau}F_{2,\ttau}\|_2+
\fr{1}{2\ttau}\|  (\nabla\H_{0,\ttau}'\cdot\nabla\etai)^{\w}  \e^{\i\ti\om(D_{\wt{x}}) \ttau}F_{2,\ttau}\|_2+O(\ttau^{-2})\\
&=\|  (\H_{\ttau}'\etai)^{\w} F_{2,\ttau}\|_2+
\fr{1}{2\ttau}\|  (\nabla_{x}\H_{\ttau}'\cdot \nabla\etai)^{\w} F_{2,\ttau}\|_2+O(\ttau^{-2}),\non
\eeqa
where $\H'(\ti x,\ti\xi):=\H'_0(\ti x-\nabla\ti\om(\ti\xi))$. As shown in Lemma~\ref{convexity}, $\pi_x\supp( \H' \etai)\cap D=\emptyset$ for $\de$ sufficiently small.  Thus for such $\de$  we can choose $G\in C_0^{\infty}(\real^{nd})$ which is equal to one on $\pi_x\supp( \H' \etai)$, supported in $(\pi_x\supp( \H' \etai))^{\de}$ and s.t.  $\supp\, G\cap D=\emptyset$.
Then, making use of Lemma~\ref{second-pseudo}, 
\beqa
\|  (\H_{\ttau}'\etai)^{\w} F_{2,\ttau}\|_2=\|  (\H_{\ttau}'\etai)^{\w}G_{\ttau}(\ti x) F_{2,\ttau}\|_2+O(\ttau^{-\infty})=O(\ttau^{-\infty}),
\eeqa
where in the last step we exploited  almost locality and the fact that $F_{2,\ttau}$ contains a commutator. As the term involving  $\nabla_{x}\H_{\ttau}'$ can be treated analogously, this concludes the proof. \qed\\
Let us add a remark about this proof. Since Lemma~\ref{decay-of-terms} gives $|F(t)|\leq C_N t^{-N}$, it may seem possible to compensate the polynomial growth with $t$ of the bound in (\ref{x}) and conclude more directly that (\ref{sum}) tends to zero. Unfortunately, this strategy does not seem to work: with the resulting dependence of $C_N$ on $\al, \be$  we were not able to control the sum in (\ref{sum}). Therefore, we followed a different route above.

\appendix
\section{Some auxiliary lemmas} \label{AppendixA}
\setcounter{equation}{0}
\bel\label{momentum-cons-lemma} Let $B_1, \ldots B_n\in \mcL_0$, $\Psi\in\hil_{\cp}$ and $F_t(x_1,\ldots, x_n):=\lan \Om, B_1(t,x_1)\ldots B_n(t,x_n)\Psi\ran$.
Then $F_t\in L^2(\Ga^{n})$ and for any $\wh{\chii}_i\in C^{\infty}(\wh\Ga)_{\real}$ which are equal to one on 
$\pi_{p}(\suppA_{B^*_i}\tau)$
\beqa
F_t=\wh\chii_1(D_{x_1})\ldots\wh\chii_n(D_{x_n})F_t. 
\eeqa
\eel
\proof It follows from (\ref{B-bound}) that $F_t\in  
L^2(\Ga^{n})$. By the same token, for $B\in\mcL_0$ and $\Phi,\Psi\in \hil_{\cp}$ 
we have $\lan \Phi, B(t,\,\cdot\,)\Psi\ran\in L^2(\Ga)$. Moreover, for $g$ as in the statement of the lemma
\beqa\bes \label{xx}
\lan \Phi, B(t,x)\Psi\ran&=\lan \tau^{(d)}_{g}(B^*)(t,x)\Phi, \Psi\ran\\
&=(2\pi)^{-d/2}\int_{\Ga} d\mu(y)\, \ov{\chii(y)} \lan B^*(t,x+y)\Phi, \Psi\ran\\
&=(2\pi)^{-d/2}\int_{\Ga} d\mu(y)\, \ov{\chii(y)}\e^{\i D_x\cdot  y} \lan\Phi, B(t,x) \Psi\ran=\wh\chii(D_x)\lan \Phi, B(t,x)\Psi\ran.
\ees\eeqa
By iterating this argument we conclude the proof. \qed\\
The following lemma is a lattice variant of a result from \cite{Bos00}. 
\bel\label{Bostelmann} Let $\etai\in C^{\infty}(\wh\Ga)_{\real}$, $\H_0\in C_0^{\infty}(\real^{nd})$. Then
 \beqa
\etai(D_{\ti x})\H_{0,t}(\ti x) \etai(D_{\ti x})=
\sum_{\al,\be\in \nat_0^{nd}}
\fr{ \wt\H_{0,t}^{\al,\be} } {\al!\be!} |\etai(\ti k)\ti k^{\al}\ran \lan \etai(\ti k)\ti k^{\be}|+O(t^{-\infty}),
\label{expansion-one-one}
\eeqa
where 
\beqa
\wt\H_{0,t}^{\al,\be}:=(2\pi)^{-\fr{nd}{2}} (-1)^{|\al|}
t^{|\al|+|\be|+nd} \sum_{\ell\in \Ga^n_1 }\wt{\H}_0^{(\al+\be)}(2\pi\ell t),
\eeqa
$\wt \H_0^{(\al+\be)}$ denotes the derivative of  the Fourier transform of $\H_0$ (on $S(\real^{nd})$) and $\Ga^n_1$ is a sufficiently large finite subset
of $\Ga^n$ containing zero (depending only on $n,d$).
\eel
\proof   As in this proof we have to consider Fourier transforms both on $S(\real^{nd})$ and $S(\Ga^n)$, we will
 treat $\wh\Ga^n=]-\pi,\pi]^{nd}$ as a subset of $\real^{nd}$. Let us consider the kernel of the operator on the 
l.h.s. of (\ref{expansion-one-one}) in momentum space
\beqa
(\etai(D_{\ti x})\H_{0,t}(\ti x) \etai(D_{\ti x}))(\ti k_1,\ti k_2)=(2\pi)^{-nd/2}t^{nd}\etai(\ti k_2) 
\sum_{\ell\in \Ga^n} \wt{\H}_0(t(\ti k_2-\ti k_1+2\pi\ell))\etai(\ti k_1),
\eeqa
where we made use of property (\ref{Dirichlet}) below. We decompose $\Ga^n=\Ga^n_1+\Ga^n_2$ s.t. 
$\Ga^n_1$ is finite and for  $\ell\in \Ga^n_2$ we have
$|\ti k_2-\ti k_1+2\pi\ell|\geq 1$ for all $\ti k_1,\ti k_2\in \wh\Ga^n$. We analyse the corresponding
kernels:
\beqa
(\etai(D_{\ti x})\H_{0,t}(\ti x) \etai(D_{\ti x}))_i(\ti k_1,\ti k_2):=(2\pi)^{-nd/2}t^{nd}\etai(\ti k_2) 
\sum_{\ell\in \Ga^n_i} \wt{\H}_0(t(\ti k_2-\ti k_1+2\pi\ell))\etai(\ti k_1).
\eeqa
We note that due to the rapid decay of $\wt\H_0$
\beqa
|(\etai(D_{\ti x})\H_{0,t}(\ti x) \etai(D_{\ti x}))_2(\ti k_1,\ti k_2)|
\leq (2\pi)^{-nd/2}t^{nd} \fr{C_{2N}}{t^N} \sum_{\ell\in \Ga^n}   \fr{1}{(c'+| 2\pi\ell |)^N  },
\eeqa
thus by Lemma~\ref{Schur} the corresponding operator gives the error term in (\ref{expansion-one-one}).

Now we analyse the leading term. We write for $\ell\in \Ga^n_1$, exploiting analyticity of $\wt\H_0$,
\beqa\bes
\wt{\H}_0(t(\ti k_2-\ti k_1+2\pi\ell))&=\sum_{\ka\in \nat_0^{nd}} \fr{t^{|\ka|}\wt{\H}_0^{(\ka)}(2\pi\ell t) }{\ka!}(\ti k_2-\ti k_1)^{\ka}\\
&=\sum_{\al,\be\in \nat_0^{nd}} (-1)^{|\al|}t^{|\al|+|\be|}\fr{\wt{\H}_0^{(\al+\be)}(2\pi\ell t) }{\al!\be!}\ti k_1^{\al}\ti k_2^{\be},
\ees\eeqa
which concludes the proof. \qed
\bel\label{HR-states} Let $\wh\chii_i$, $i=1,\ldots, n$, be as in Lemma~\ref{momentum-cons-lemma} and supported in the interior
of $]-\pi,\pi]^d$.
 Then 
\beqa
\lan \Psi,a^*_{\un B}\big(|\Om\ran\otimes \e^{-\i t\ti\om(D_{\ti x})} |\etai(\ti k)\ti k^{\be}\ran)= \lan\Psi,B_{1,t}^*(g^{\be_1}_{1,t})\ldots B_{n,t}^*(g^{\be_n}_{n,t})\Om\ran,
\label{scattering-states-a-star}
\eeqa
where
\beqa
g^{\be_i}_{i,t}(x)=(2\pi)^{-d/2} \int_{\wh\Ga} dk\, 
\e^{-\i\om(k)t+\i k\cdot x}\wh\chii_{i}(k)k^{\be_i}, \quad i=1,\ldots,n,
\label{g-support}
\eeqa
and $\be=(\be_1,\ldots, \be_n)$, $\be_i\in \nat_0^{d}$.  
\eel
\proof The observation that $\lan \ti x |\e^{-\i t\ti\om(D_{\ti x})} |\etai(\ti k)\ti k^{\be}\ran=g_{1,t}^{\be_1}(x_1) \ldots g_{n,t}^{\be_n}(x_n)$ proves (\ref{scattering-states-a-star}). Due to the assumption that $\wh\chii_i$ are supported in the interior of $]-\pi,\pi]^d$, we have that
$\wh\chii_{i}(k)k^{\be_i}\in C^{\infty}(\wh\Ga)$ and (\ref{g-support}) is a wave packet in the sense of definition~(\ref{wave-packet-def}).
 \qed
\bel\label{decay-of-terms}  Let $g_{i,\ttau}^{\be_i}$ be as in Lemma~\ref{HR-states} and s.t. $\vel(\supp\,\wh\chii_i)$ are disjoint.
Then, for  any compact set $\De$ we have 
\beqa
&\|B_{i,\ttau}^*(g_{i,\ttau}^{\be_i})P(\De)\|\leq c^{|\be_i|} , \label{simple-bound-B}\\
&\|P(\De) [B_{1,\ttau}^*(g_{1,\ttau}^{\be_1}),  
B_{2,\ttau}^*(g_{2,\ttau}^{\be_2}) ]P(\De)\|\leq c_N\ttau^{-N} 
c^{|\be_1|+|\be_2|},
\eeqa
where $c$ is independent of $\be_i$ and $\ttau$.
The bound remains true if $B_{1,\ttau}^*$ is replaced with $\pa_{\ttau}(B^*_{1,\ttau}(g_{1,\ttau}^{\be_1}))$. 
\eel
\proof By Theorem~\ref{harmonic-theorem}, compact energy-momentum transfer of $B_i$ and the Cauchy-Schwarz inequality we immediately get
\beqa
\|B_{i,\ttau}^*(g_{i,\ttau}^{\be_i})P(\De)\|\leq c'\|g_i^{\be_i}\|_2\leq c^{|\be_i|},
\eeqa
where in the last step we used that $\wh\Ga$ is a compact set.

Now let $\chi_i\in C_0^{\infty}(\real^d)$ be approximate characteristic functions of the velocity supports $\vel(\supp\, \wh g_i)$
and $\chi_i':=1-\chi_i$. Clearly, we can choose $\chi_i$ 
with disjoint supports. Thus setting $\chi_{i,\ttau}(x):=\chi_i(x/\ttau)$ we have by almost locality of $B_i$
\beqa
\|[B_{1,\ttau}^*(\chi_{1,\ttau} g_{1,\ttau}^{\be_1}),  
B_{2,\ttau}^*(\chi_{2,\ttau}g_{2,\ttau}^{\be_2}) ]\|\leq c_N\ttau^{-N}  \|g_{1}^{\be_1}\|_2\|g_{2}^{\be_2}\|_2\leq c_N \ttau^{-N}c^{|\be_1|+|\be_2|} .
\eeqa
Now  terms involving $\chi'$ can be estimated as follows using (\ref{simple-bound-B})
\beqa
\|[B_{1,\ttau}^*(\chi'_{1,\ttau} g_{1,\ttau}^{\be_1}),  
B_{2,\ttau}^*(\chi_{2,\ttau}g_{2,\ttau}^{\be_2}) ]\one_{\De}(U) \|\leq C\|\chi'_{1,\ttau} g_{1,\ttau}^{\be_1}\|_2\|g_{2,\ttau}^{\be_2}\|_2.
\eeqa
By a standard application of the non-stationary phase method (Theorem XI.14 of \cite{RS3}) we obtain
\beqa
\|\chi'_{1,\ttau} g_{1,\ttau}^{\be_1}\|_2\leq c_N \ttau^{-N}c^{|\be_1|}  
\eeqa
which concludes the proof. \qed
\bel\label{convexity} With definitions as in the proof of Theorem~\ref{Main-result}, $\pi_x\supp( \H' \etai)\cap D=\emptyset$ for $\de$
sufficiently small.
\eel
\proof Recall that
\beqa\bes
 &\H'(\ti x,\ti\xi):=\H'_0(\ti x-\nabla\ti\om(\ti\xi)), \quad \supp\,\H'_0\subset  (\pi_{\ti x}\, \supp\, (\H_1\etai') )^{\de,\cv,\de}, \\
&\H_{1}(\ti x, \ti\xi):=\H(\ti x+\nabla\ti\om(\ti\xi)), \quad
\ \ \supp\, \H^{\cv} \cap D=\emptyset. 
\ees\eeqa
Making use of the facts that for  $X,Y, Z\subset \real^{\ell}$ we have $(X+Y)^{\cv}=X^{\cv}+Y^{\cv}$, $(X+Y)^{\de}\subset X^{\de}+Y^{\de}$,  
$Z^{\de,\cv}=Z^{\cv,\de}$, we obtain 
\beqa\bes \label{conv-comp}
\pi_x\supp( \H' \etai)&\subset \supp\, \H'_0+\vel(\supp\,\wh g)\\
&\subset (\pi_{\ti x}\, \supp\, (\H_1\etai') )^{\de,\cv,\de}+\vel(\supp\,\wh g)\\
&\subset (\supp\, \H-\vel(\supp\,\wh g')   )^{\de,\cv,\de}+\vel(\supp\,\wh g)\\
&\subset (\supp\, \H)^{\cv,2\de}-(\vel(\supp\,\wh g'))^{\cv,2\de}+\vel(\supp\,\wh g)\\
&\subset (\supp\, \H)^{\cv, R(\de)},
\ees\eeqa
where $R(\de)\to 0$ as $\de\to 0$. Here we denoted
\beqa
\vel(\supp\,\wh{g}):=\{\nabla\ti\Si(\ti \xi)\,|\, \ti\xi\in \supp\,\wh g\,\}=\vel(\supp\,\wh{g}_1)
\times\cdots\times \vel(\supp\, \wh{g}_n),
\eeqa
and analogously for $\vel(\supp\,\wh{g}')$. In the last step of (\ref{conv-comp})
we used that $\supp\,\wh g_i$, $\supp\,\wh g'_i$
are contained in some balls $\mathcal{B}(k_i,3\de)$ and therefore, by
continuity of $\nabla\Si$, the velocity supports $\vel(\supp\,\wh g_i), \vel(\supp\,\wh g'_i)$
are contained in some balls $\mathcal{B}( v_i ,r(\de))$, where $r(\de)\to 0$
with $\de\to 0$. This gives
\beqa\bes
\vel(\supp\,\wh{g})-\vel(\supp\,\wh{g}')^{\cv,2\de}&\subset \mathcal{B}( \ti v, \sqrt{n}r(\de))- \mathcal{B}( \ti v, \sqrt{n}r(\de)+2\de)\\
&\subset \mathcal{B}(0,2\sqrt{n}r(\de)+4\de),
\ees\eeqa
and therefore the last line of (\ref{conv-comp}). Since $(\supp\, \H)^{\cv}\cap D=\emptyset$, we conclude that
$\pi_x\supp( \H' \etai)\cap D=\emptyset$ for
sufficiently small $\de$. \qed

\section{Pseudo-differential calculus on a lattice} \label{pseudodifferential} 
\setcounter{equation}{0}

A systematic exposition of pseudo-differential calculus on a lattice can be found e.g. in Appendix B of \cite{GN98}.
However, for the reader's convenience, we derive the particular properties needed in this paper, as they are not easy to extract from the literature.
Lemmas~\ref{trivial-pseudo}, \ref{second-pseudo} are well known in the continuum case and the proofs
are obvious generalizations, so we can be brief. Lemma~\ref{free-evolution-x} is less standard and the lattice 
causes some complications, so we give more details.  
We start with some definitions: 
\begin{enumerate}
\item We say that $a\in S(\real^{\m}\times \wh\Ga)$ if $a\in C^{\infty}(\real^{\m}\times \wh\Ga)$ and $\sup_{x\in\real^{\m}, \xi\in \wh\Ga} |x^{\al}\pa^{\be}_x\pa^{\ga}_{\xi}a(x,\xi)|<\infty.$ 
\item We define a family of bounded self-adjoint operators $D_x^i$, $i=1,\ldots,d$, on $L^2(\Ga)$ by their action in momentum space
\beqa
\widehat{D_{x}^i\phi}(p)=p^i\wh{\phi}(p),
\eeqa
where 
we use the parametrisation
$\wh\Ga=]-\pi,\pi]^{\m}$. It immediately follows that $(\e^{\i D_x\cdot y}\phi)(x)=\phi(x+y)$,
where $\phi\in L^2(\Ga)$, $y\in \Ga$.
\end{enumerate}
Now we define the Weyl quantization of a symbol $a\in S(\real^{\m}\times \wh\Ga)$: For $\phi,\psi\in L^2(\Ga)$ we write
\beqa\bes
(\phi,a^{\w}\psi)&=(2\pi)^{-\m}\int_{\Ga} d\mu(x)\int_{\Ga}d\mu(y)\int_{\wh\Ga} d\xi \, a\left(\fr{x+y}{2},\xi\right)\bar\phi(x)\psi(y)
\e^{\i (x-y)\cdot \xi}\\
&=(2\pi)^{-\m/2}\int_{\Ga} d\mu(x)\int_{\Ga}d\mu(y) \ \widecheck{a}\left(\fr{x+y}{2},x-y\right)\bar\phi(x)\psi(y), \label{kernel-lattice}
\ees\eeqa
where check denotes here the inverse Fourier transform in the second variable.
\newcommand{\g}{\wh g}
\bel\label{trivial-pseudo} Let $h\in S(\real^{\m})$, $g\in S(\Ga)$. Then, with $h_t(x):=h(x/t)$,
\beqa
h_t(x) \g(D_{x})=(h_t \g)^{\w}+\fr{\i}{2t}((\nabla h)_t\cdot \nabla \g)^{\w}+O(t^{-2}),
\eeqa
where the l.h.s. is defined via functional calculus.
\eel
\begin{proof} By formula~(\ref{kernel-lattice}) and standard considerations we determine the kernels of $(h_t \g)^{\w}$ and $h_t(x) \g(D_{x})$
\beqa\bes
(h_t \g)^{\w}(x,y)&=(2\pi)^{-\m/2}h\bigg(\fr{x+y}{2t}\bigg) g(x-y), \\
(h_t(x) \g(D_{x}))(x,y)&=(2\pi)^{-\m/2}h(x/t)g(x-y).
\ees\eeqa
Next we write
\beqa
(h_t(x) g(D_{x})\psi)(x,y)=(2\pi)^{-\m/2}h((x+y)/(2t)+(x-y)/(2t)   ) g(x-y).
\eeqa
We set $z:=(x+y)/2$, $w:=(x-y)$,  denote
\beqa
K_s(x,y):=h(z/t+sw/(2t)) g(w)
\eeqa
and expand $K_s$ into the Taylor series w.r.t. $s$:
\beqa\bes
K_1(x,y)&=h(z/t) g(w)+\fr{1}{2t}(\nabla h)(z/t)\cdot w g(w)\\
&+\fr{1}{(2t)^2}\int_0^1ds'(1-s')\pa_i\pa_j h(z/t+s'w/(2t))  w^{i} w^j  g(w). \label{kernel-expansion}
\ees\eeqa
It is clear that the first two terms on the r.h.s. of (\ref{kernel-expansion}) are kernels of 
$(h_t \g)^{\w}$ and $\fr{\i}{2t}((\nabla h)_t\cdot \nabla \g)^{\w}$. The norm od the last term 
is estimated using Lemma~\ref{Schur} below. \end{proof}
\bel\label{second-pseudo} Let $a\in S(\real^{\m}\times \wh\Ga)$ and $h \in S(\real^{\m})$ be s.t. $h=1$ on $\pi_x\supp\, a$. Then
\beqa
(1-h_t(x))a_t^{\w}=O(t^{-\infty}).
\eeqa
\eel
\proof  Taylor expansion of kernels, analogous to the proof of Lemma~\ref{trivial-pseudo}. \qed
\bel\label{free-evolution-x} Let $a\in  S(\real^{\m}\times \wh\Ga)$ and $\om\in C^{\infty}(\wh\Ga)$. Then
\beqa
\e^{\i\om(D_x) t }a_t^\w\e^{-\i\om(D_x) t }=  a_{1,t}^{\w}+O(t^{-2}),  \label{free-evolution}
\eeqa 
where $a_{1}(x,\xi)=a(x+\nabla\om(\xi),\xi)$. 
\eel
\proof  
As in this proof we have to consider Fourier transforms both on $S(\real^{\m})$ and $S(\Ga)$, it will be
at times convenient to treat $\wh\Ga=]-\pi,\pi]^{\m}$ as a subset of $\real^{\m}$ and functions on $\wh\Ga$
as periodic functions on $\real^{\m}$.

For $\eps\in \{0,1\}$ we define the operator $K_{\eps,t}$ on $L^2(\Ga)$ given by the kernel
\beqa\bes
K_{\eps,t}(x,y)&=(2\pi)^{-\m}\int_{\wh\Ga} d\xi\, a\bigg(\fr{x+y}{2t}+\eps\nabla\om(\xi), \xi\bigg)\e^{\i \xi\cdot (x-y) } \\
&=(2\pi)^{-3\m/2}t^{\m}\int_{\wh\Ga} d\xi\,\int_{\real^n} d\xi' \,\wt{a} (t\xi', \xi)\e^{\i \eps t \xi' \cdot  \nabla\om(\xi)}\e^{\i \xi \cdot (x-y) }\e^{\i \h \xi'\cdot (x+y)},
\ees\eeqa
where tilde denotes the Fourier transform (on $S(\real^{\m})$) in the first variable. Now using the Parseval theorem and the fact that for $p\in \real^{\m}$
\beqa
(2\pi)^{-\m}\int_{\Ga} d\mu(x)\, \e^{\i p\cdot x}=\sum_{\ell\in \mathbb{Z}^{\m}}\de(p-2\pi \ell), \label{Dirichlet}
\eeqa 
as an equality on $S(\real^\m)$,
we obtain the kernel of $K_{\eps,t}$ in momentum space
\beqa\bes
K_{\eps,t}(p,q)&=(2\pi)^{-\m} \int_{\Ga^2}d\mu(x) d\mu(y)\,  \e^{-\i p\cdot x+\i q\cdot y } K_{\eps,t}(x,y)\\
&=(2\pi)^{-\m/2} t^{\m}\sum_{\ell_1,\ell_2\in \Ga} \chi\big(  \xi_{\ell}  \in \wh{\Ga} )\e^{\i \eps t \xi'_{\ell}\cdot  \nabla\om(\xi_{\ell}) } 
\wt{a} (t \xi'_{\ell} , \xi_{\ell} ),
\ees\eeqa
where $p,q\in \wh\Ga$, $\chi$ is the characteristic function  and in the last step we  set for $\ell=(\ell_1,\ell_2)$:
\beqa
\xi_{\ell}:=(p+q)/2+\pi(\ell_2-\ell_1), \quad
\xi'_{\ell}:=p-q-2\pi(\ell_1+\ell_2).
\eeqa
We will study the kernel of the operator on the l.h.s. of (\ref{free-evolution}), which is $\e^{\i t \om(p)}K_{0,t}(p,q)  \e^{-\i t \om(q)}$, and show that its leading part gives the kernel of the first term on the r.h.s. of (\ref{free-evolution}), which is $K_{1,t}(p,q)$.
Using periodicity of $\om$, we obtain for any $\ell$ 
\beqa
& \om(p)
=\om\bigg(\xi_{\ell}+\h\xi'_{\ell} \bigg),\quad
 \om(q)
=\om\bigg(\xi_{\ell}-\h\xi'_{\ell} \bigg).
\eeqa
Consequently, we can write
\beqa
 \e^{\i t \om(p)}K_{0,t}(p,q)  \e^{-\i t \om(q)}
=(2\pi)^{-\m/2} t^{\m}\sum_{\ell_1,\ell_2\in \Ga} \chi\big( \xi_{\ell}  \in \wh{\Ga} )
\wt{a} (t(\xi'_{\ell}), \xi_{\ell} )\e^{\i t(\om(\xi_{\ell}+\h\xi'_{\ell} )-\om(\xi_{\ell}-\h\xi'_{\ell}  )) }. \label{rotated-kernels}
\eeqa
We define the function of $\la\in \real$
\beqa
f_{\xi_{\ell}, \xi_{\ell}' }(\la):=\om(\xi_{\ell}+\la\xi'_{\ell}/2)-\om(\xi_{\ell}-\la\xi'_{\ell}/2),
\eeqa
and consider the Taylor expansion
\beqa\bes
& f_{\xi_{\ell}, \xi_{\ell'}}(1)=\nabla\om(\xi_{\ell})\cdot \xi'_{\ell}+R_{\xi_{\ell}, \xi_{\ell}' }, \\
& R_{\xi_{\ell}, \xi_{\ell}'}:=\sum_{\substack{\al\in \nat_0^d  \\ |\al|=3}}  \int_0^1 d\la' \, (1-\la')^2  \om_{\al,\la'}(\xi_{\ell},\xi'_{\ell})    (\xi'_{\ell})^{\al},  \\
& \om_{\al,\la'}(\xi_{\ell},\xi'_{\ell}):=\fr{1}{2!2^3}((\pa^{\al}\om)(\xi_{\ell}+\la'\xi'_{\ell}/2)-(\pa^{\al}\om)(\xi_{\ell}-\la'\xi_{\ell}'/2)).
\ees\eeqa
Here we exploited that $f_{\xi_{\ell}, \xi_{\ell}' }(0)=f_{\xi_{\ell}, \xi_{\ell}' }^{(2)}(0)=0$. (Due to this fact
the error term in (\ref{free-evolution}) is only $O(t^{-2})$). 
The above considerations give
\beqa
\e^{\i t(\om(p)-\om(q))}=\e^{\i t\nabla\om(\xi_{\ell})\cdot \xi'_{\ell}} \e^{\i t R_{\xi_{\ell}, \xi_{\ell}'   }  }
=\e^{\i t\nabla\om(\xi_{\ell})\cdot \xi'_{\ell}} +\e^{\i t\nabla\om(\xi_{\ell})\cdot \xi'_{\ell}} \int_0^{1} d\la'' \e^{\i t\la'' R_{ \xi_{\ell}, \xi_{\ell}' } } 
\i tR_{\xi_{\ell}, \xi_{\ell}' }. \label{second-Taylor}
\eeqa
First term on the r.h.s. of (\ref{second-Taylor}), substituted to (\ref{rotated-kernels}), gives indeed $K_{1,t}(p,q)$. 
We estimate the remainder using 
Lemma~\ref{Schur}. Thus we are  interested in the norm of the operator $r$ given by the kernel
\beqa
r(p,q):=(2\pi)^{-\m/2}  t^{\m}\sum_{\ell_1,\ell_2}  \wt a\left(t\xi'_{\ell},\xi_{\ell}\right)\e^{\i (t\xi'_{\ell})\cdot\nabla\om(\xi_{\ell}) } \chi\big( \xi_{\ell}  \in \wh{\Ga} ) 
\int_0^{1} d\la'' \e^{\i t\la'' R_{\xi_{\ell}, \xi_{\ell}' } } \i tR_{\xi_{\ell}, \xi_{\ell}' }.
\eeqa
We note the estimate:
\beqa
|r(p,q)|\leq ct^{\m+1}\sum_{\ell_1,\ell_2}  |\wt{a}\left(t \xi'_{\ell}, \xi_{\ell}\right)|    |\xi'_{\ell}|^{3} 
\chi\big( \xi_{\ell}  \in \wh{\Ga} ).
\eeqa
We verify the assumptions of Lemma~\ref{Schur}: We set $\ell_{\pm}=\ell_2\pm \ell_1$ and compute
\beqa\bes
\sup_{p\in \wh\Ga}\int_{\wh\Ga} dq\, |r(p,q)|&\leq ct^{\m+1}\sup_{p\in \wh\Ga}\sum_{\ell_1,\ell_2}\int_{\wh\Ga} dq |\wt a\left(t \xi'_{\ell},\xi_{\ell}\right)|\,|\xi'_{\ell}|^{3}\chi\big( \xi_{\ell}  \in \wh{\Ga} ) \\
&\leq  ct^{\m+1}\sup_{p\in \wh\Ga}\sum_{\ell_{+},\ell_-}\int_{\wh\Ga} dq |\wt a\left(t( p-q-2\pi\ell_{+} ), (p+q)/2+\pi\ell_- \right)| \\ & \phantom{444444444444444444}\times
|p-q-2\pi \ell_+ |^{3}\chi(  (p+q)/2+\pi\ell_-  )   \\
&\leq c_Nt^{\m+1}\sup_{p\in \wh\Ga}\sum_{\ell_{+}}\int_{\wh\Ga} dq \,\lan t( p-q-2\pi \ell_{+}) \ran^{-N}
|p-q-2\pi \ell_+ |^{3}\\
&\leq c_Nt^{\m+1} \int_{\real^d} dq\, \lan tq \ran^{-N}
|q |^{3}
\leq c_Nt^{-2}  \int_{\real^d} dq\, \lan q \ran^{-N}
|q |^{3}.
\ees\eeqa
Here in the third step we used the rapid decay of $\wt a$ in the first variable and fact that $\chi((p+q)/2+\pi\ell_-)$ together with the condition $p,q\in \wh\Ga$ restrict the range
of $\ell_-$ to a bounded set. In the fourth step we used that the union of $\wh\Ga+2\pi\ell_+$ over all $\ell_+\in \mathbb{Z}^d$ is $\real^d$. The second bound from Lemma~\ref{Schur} is verified analogously. \qed\\ 
The following fact is known as the Schur lemma. Its proof is elementary  (see e.g. \cite{DG97}).
\bel \label{Schur} Let $Y,Y'$ be spaces with measures $d\mu$, $d\mu'$. Let $k(\,\cdot\,, \, \cdot \,)$
be a measurable function on $Y\times Y'$ s.t.
\beqa
\mathrm{essup}_Y\int |k(y,y')|d\mu'\leq C, \quad \mathrm{essup}_{Y'}\int |k(y,y')|d\mu\leq C'.
\eeqa
Then the operator $K: L^2(Y,d\mu')\to L^2(Y, d\mu)$ given by the kernel $k$ is bounded and $\|K\|\leq (CC')^{1/2}$.
\eel

\end{document}